\documentstyle[fleqn,epsfig]{article}

\newcommand{\etal}{{\em et al.}\bs}

\newcommand{\hs}[1]{\hspace{#1}}
\newcommand{\bs}{\hspace{5.0pt}}

\newcommand{\bl}{\hspace{3.0pt}}

\newcommand{\tnn}{\mbox{$\tau_{n\overline{n}}\ $}}

\newcommand{\TA}{\mbox{$T_{A}$}\bs}
\newcommand{\TAn}{\mbox{$T_{A}$}}

\newcommand{\TR}{\mbox{$T_{R}$}\bs}
\newcommand{\TRn}{\mbox{$T_{R}$}}

\newcommand{\mulc}[3]{\multicolumn{#1}{#2}{#3}}

\newcommand{\mb}[1]{\mbox{#1}}

\newcommand{\frejus}{Fr\'{e}jus\hs{0.2cm}}
\newcommand{\frejusn}{Fr\'{e}jus}

\newcommand{\evis}{\mbox{$E_{vis}$}\bs}
\newcommand{\evisn}{\mbox{$E_{vis}$}}

\newcommand{\n}{$\overline{\mb{\rm n}}$\hs{0.2cm}}       
\newcommand{\nn}{${\rm n}\overline{\rm n}$\hs{0.2cm}}    
\newcommand{\nN}{\mbox{$\rm {\overline{n}N}$}\hs{0.2cm}} 
          
\newcommand{\numu}{\mbox{$\nu_{\mu}$}}                   
\newcommand{\nue}{\mbox{$\nu_{e}$}}                      

\newcommand{\del}{\delta}				 

\title{
\begin{flushright}
\vskip -50pt
{\normalsize
PDK-788 \\
TUHEP-02-01 \\
\vskip -10pt
May 2002}
\end{flushright}
\vskip 50pt
{\bf Search for Neutron-Antineutron Oscillations \\
Using Multiprong Events in Soudan 2} 
\vskip 30pt
}

\author{
J.~Chung$^5$,
W.W.M.~Allison$^3$, G.J.~Alner$^4$, D.S.~Ayres$^1$, W.L.~Barrett$^6$,\\ 
P.M.~Border$^2$, J.H.~Cobb$^3$, H.~Courant$^2$, D.M.~Demuth$^2$\\
T.H.~Fields$^1$, H.R.~Gallagher$^5$, M.C.~Goodman$^1$,
R.~Gran$^2$, \\ T.~Joffe-Minor$^1$, T.~Kafka$^5$, S.M.S.~Kasahara$^2$, 
P.J.~Litchfield$^4$, \\ W.A.~Mann$^5$, M.L.~Marshak$^2$,
R.H.~Milburn$^5$, W.H.~Miller$^2$,  L.~Mualem$^2$, \\ A.~Napier$^5$,
W.P.~Oliver$^5$, G.F.~Pearce$^4$, E.A.~Peterson$^2$, D.A.~Petyt$^4$,\\
K.~Ruddick$^2$, M.~Sanchez$^5$, J.~Schneps$^5$, A.~Sousa$^5$,
\\ B.~Speakman$^2$, J.L.~Thron$^1$, S.P.~Wakely$^2$, N.~West$^3$\\
\\
$^1${\it Argonne National Laboratory, Argonne, IL 60439}\\
$^2${\it University of Minnesota, Minneapolis, MN 55455}\\
$^3${\it Department of Physics, University of Oxford, Oxford OX1 3RH, UK}\\
$^4${\it Rutherford Appleton Laboratory, Chilton, Didcot, Oxfordshire
 OX11 0QX, UK}\\
$^5${\it Tufts University, Medford, MA 02155}\\
$^6${\it Western Washington University, Bellingham, WA 98225}\\
}

\begin{document}


\maketitle 
\thispagestyle{empty}

\begin{abstract}
\normalsize
We have searched for neutron-antineutron oscillations using 
the 5.56 fiducial kiloton-year exposure of the Soudan 2 
iron tracking calorimeter.  We require candidate \nn occurrences   
to have $\ge$ 4 prongs (tracks and showers) and to have kinematics 
compatible with \nN annihilation within a nucleus. 
We observe five candidate events,
with an estimated background from atmospheric neutrino and cosmic ray
induced events of 4.5 $\pm$ 1.2 events.
Previous experiments with smaller exposures observed no candidates, 
with estimated background rates similar to this experiment. 
We set a lifetime lower limit at 90\% CL for
the \nn oscillation time in iron:
$\TAn(Fe) >  7.2 \times 10^{31}$~years. 
The corresponding lower limit for oscillation of free neutrons
is $\tnn > 1.3 \times 10^{8}$~seconds.


\end{abstract}
\eject

\large

\section{Introduction}

\subsection{Neutron-antineutron oscillations}

     An intriguing variation on the grand unification theme that nucleons
are likely to be unstable is the proposal that 
neutrons can oscillate into antineutrons.  
Neutron-antineutron oscillations were first predicted in 1970 by V. A. Kuzmin 
in a model intended as a realization of requirements given earlier  
by A.D. Sakharov for evolution of the universe to net baryon asymmetry
\cite{{Kuzmin_70},{Sakharov_67}}.  Subsequently \nn
oscillations emerged as a predicted reaction in certain grand unification
theories \cite{GUTS_non-SUSY}.
More recently it has been shown that \nn oscillations can occur 
in a large class of supersymmetric
$SU(2)_{L} \times SU(2)_{R} \times SU(4)_{c}$
models \cite{GUTS_SUSY}.  In such models 
the dominant baryon number violating process is a 
${\it \Delta} B = -2, {\it \Delta} L = 0$ nucleon transition 
(e.g. \nn oscillations or p + n $\rightarrow$ pions)
rather than a ${\it \Delta} B = -1, {\it \Delta}L = -1
$ nucleon-antilepton transition 
(e.g.  p $\rightarrow e^{+} \pi^{0}$ or p $\rightarrow \overline{\nu} K^{+}$). 
Neutron-antineutron oscillations have also been indicated as viable
by recent GUT models which invoke the existence 
of extra spacetime dimensions \cite{GUTS_higherD}.

    If indeed a neutron can evolve into an antineutron, 
the experimental signatures for the metamorphosis 
should be distinctive.  The resulting antineutron will 
annihilate with a baryon of the surrounding environment, producing
multiple mesons ($B=0$) whose visible energy and net momentum
are approximately those of two nucleon masses having nuclear Fermi motion.

   From the phenomenology of neutron-antineutron oscillations it can 
be shown \cite{ORNL-6910_Alberico} that the oscillation time $T_A$
of a neutron bound within a nucleus of atomic mass $A$ is related to 
the neutron oscillation time in vacuum \tnn  ~according to

\begin{equation}
T_A = (\tnn)^2 \cdot T_R .
\label{equ:bound_neutron_time}
\end{equation}
 
\noindent Here $T_R$, which has units of inverse time, 
is the suppression factor representing the effect 
of the nuclear environment which substantially prolongs the effective
oscillation time.

   Detailed calculations of the suppression factor $T_R$ for parent nuclei 
of experimental interest, including deuterium, oxygen, argon, and iron, 
have been reported in literature.
The calculations utilize phenomenological frameworks provided by  
nuclear potential theory 
\cite{{Dover_83_85_89},{Huefner_98}}, and by S-matrix theory
\cite{{Chetyrkin_81},{Alberico_82_84},{Alberico_91},{Kondratyuk_96}}.  
In the analysis of Dover, Gal, and Richard
\cite{Dover_83_85_89}, 
it is proposed that neutron-antineutron oscillations 
will occur mostly in outer nuclear shells 
and near the nuclear surface.
However reservations concerning this picture
have been expressed and
in a number of calculations
the entire nuclear volume contributes to 
\nn oscillations \cite{Huefner_98}.

\subsection{ Previous experimental searches }

   Two types of experiments have been used to search 
for neutron-antineutron oscillations. 
In one approach, slow neutrons from a fission reactor
are channeled through a magnetically shielded vacuum pipe towards a
target region.  An antineutron produced during the flight
will annihilate in the target and the annihilation products are
registered by detectors surrounding the target.
Experiments of this type have been carried out at Pavia 
\cite{Bressi_89} and at Grenoble \cite{{Fidecaro_85},{Baldo-Ceolin_90_94}}.
The Grenoble reactor experiment obtained
\tnn \bs $\ge 0.86 \times 10^8$ s at 90\% confidence level (CL).  This is
the most stringent oscillation time lower limit reported to date using
free neutrons. 

   The alternate approach, used in this experiment, is to continuously
monitor neutrons bound in nuclei, usually as part of an 
ongoing nucleon decay search.  Searches of this type have been reported 
by the underground experiments Homestake \cite{Homestake_83}, 
NUSEX \cite{NUSEX_83}, KOLAR \cite{KOLAR_86}, 
IMB \cite{IMB_84}, Kamiokande \cite{KAM_86}, and Fr\'{e}jus \cite{Frejus_90}.
The searches by Kamiokande and Fr\'{e}jus 
obtained the most stringent \nn oscillation time lower limits. 

   In a search based upon a 1.11 kiloton-year (kty) exposure of the
Kamiokande-I water Cherenkov detector \cite{KAM_86}, 
no candidate \nN event was observed.  An oscillation time lower limit 
of $T_{A} >$ 4.3 $\times 10^{31}$ years at 90\% CL was set.  
Using the suppression factor $T_{R} = 1 \times 10^{23}$ s$^{-1}$
calculated by Dover {\it et al.} for oxygen \cite{Dover_83_85_89},
Kamiokande obtained an oscillation time limit for free neutrons of
$\tnn > 1.2 \times 10^8$ seconds at 90\% CL.
The Fr\'{e}jus collaboration, in a search using a 1.56 fiducial kty exposure 
of the experiment's planar iron tracking calorimeter,
also reported zero \nn oscillation candidates.
The oscillation time lower limit for $T_{A}$ in iron thereby obtained was
$6.5 \times 10^{31}$ years at 90\% CL.
Using  $T_{R} = 1.4 \times 10^{23}$ s$^{-1}$ 
as calculated by Dover {\it et al.}
for iron, Fr\'{e}jus determined the free neutron limit to be also
$\tau_{n\overline{n}}$ $> 1.2 \times 10^8$ seconds
at 90\% CL \cite{Frejus_90}.

   In the Kamiokande analysis an enhanced probability for \nn oscillations
to occur in the nuclear periphery as postulated by Dover {\it et al.} was
assumed.  In the Monte Carlo simulations of the experiment, the effect of
this assumption is to reduce distortion of the final state meson spectrum
arising from intranuclear absorption and inelastic scattering processes.
As a result experimental detection efficiencies are enhanced 
relative to expectations for the case where oscillations may occur 
throughout the entire volume of parent nuclei.  
The search reported here follows the more conservative approach 
adopted previously by \frejusn.  For our primary \nN simulation on 
which our detection efficiency is based, we assume
\nn oscillation to occur throughout the nuclear volume.  
However the effects of peripheral predominance are also described.

\section{Detecting \nn Oscillations in Soudan 2}

\subsection{Detector and data exposure}

   Soudan 2 is a 963 metric ton (770 tons fiducial) 
iron tracking calorimeter 
of honeycomb lattice geometry which operates as a slow--drift 
time projection chamber \cite{S2:NIM_A376_A381}.  The 
tracking elements are one-meter-long drift tubes
filled with an Argon--CO$_{2}$ gas mixture.   Electrons liberated by 
throughgoing charged particles drift to the tube ends under the action
of a voltage gradient applied along the tubes.  The drift charge is
registered by vertical anode wires, while horizontal
cathode pad strips register the image charges.  The third coordinate is
obtained from the drift--time.  The amount of charge measures the
deposited ionization.  The drift tubes 
are laid onto corrugated steel sheets, and the sheets are stacked to form 
1$\times$1$\times$2.5 m, 4.3 ton modules 
from which the calorimeter is assembled in building-block fashion.
Surrounding the tracking calorimeter on all sides but mounted on the 
cavern walls and well separated from calorimeter surfaces is
a 1700 m$^{2}$ active shield array of two or three layers of 
proportional tubes \cite{S2:NIM_A276}.  The shield facilitates 
identification of events which are not contained within the calorimeter.
In particular, it provides tagging of background events initiated by
cosmic-ray-induced neutrons.

   The detector is located at a depth of 2070 meters--water--equivalent 
on the 27th level of the Soudan Underground Mine State Park 
in northern Minnesota.  
The modular design enabled data taking to commence 
in April 1989 when the detector was one-quarter of its final size; 
routine operation with the fully--deployed detector got underway
in November 1993.  The fiducial (total) exposure analyzed here, 
obtained from data--taking through December 2000, 
is 5.56 (6.96) kiloton--years.
   
   Calibration of calorimeter module response
was carried out at the Rutherford ISIS spallation 
neutron facility using test beams of positive and negative 
pions, electrons, muons, and protons \cite{Wall_PRD62_00}.
Spatial resolutions for track reconstruction and for vertex
placement in anode, cathode, and drift time coordinates are of the
same scale as the drift tube radii, $\approx$ 0.7 cm.
In Soudan 2, ionizing particles having non-relativistic as well 
as relativistic momenta are imaged with $dE/dx$ sampling in a 
fine--grained honeycomb lattice geometry.
Protons can be distinguished from pions and muons 
via ionization and ranging, energetic
muons discriminated from pions via absence of secondary scatters,  
and prompt e$^{\pm}$ showers distinguished from photon showers 
on basis of proximity to primary vertices.
These event imaging capabilities offer advantages, in comparison to 
water Cherenkov detectors and to planar iron calorimeters, 
for analysis of the complicated multiprong
topologies that would arise with \nN annihilations initiated by
\nn oscillations.

\subsection{Simulation of \nN annihilation arising from \nn oscillations}

   We have developed realistic simulations of \nn oscillations yielding
\nN annihilations as they would occur in the iron nuclei which
comprise the bulk of the calorimeter mass.  
Generation of \nN events is carried out as follows:

   Momenta of the initial state antineutron and nucleon 
are assigned according to a distribution based upon a Fermi-gas model 
parameterization of quasi-elastic electron nucleus scattering \cite{Bodek_81}. 
Final state particle four-momenta were 
constructed in accordance with N-body phase space \cite{Chaffee_SAGE}.
Assignment of \nN reactions to generated events is weighted according to
cross section data for $\overline{\rm p}$p annihilation at rest 
\cite{{Backenstoss_83},{pbar_p_Refs}}.
For the purpose of channel selection, $\overline{\rm n}$n annihilations
are assigned the same cross sections as observed for 
$\overline{\rm p}$p;
$\overline{\rm n}$p is assumed to have the
same total rate as $\overline{\rm p}$p, and cross sections
were inferred from $\overline{\rm p}$p as allowed by charge conservation.  We
restricted our reaction compilation to cross sections 
exceeding 2\% of the total $\rm {\overline{N}N}$ cross section.  Consequently 
production of $\rho$, $\omega$, charged
and neutral pions and kaons is represented,
however final states with $\eta$ and $\eta'$ mesons were neglected.

   Provision was made to include intranuclear rescattering (INS)
of final state pions in the simulations. Our treatment follows 
the approach utilized previously in simulations of
atmospheric neutrinos and of nucleon decay \cite{Wall_PRD62_00}. 
Pions, either directly produced in annihilation or in
$\rho$ decays, are propagated through a model nucleus.  
The nuclear radius and density are parameters of the model; 
the radius is scaled according to $A^{1/3}$.   
Scattering in the nuclear medium is
characterized using a momentum-dependent pion interaction 
length.  The nuclear parameters were set by requiring the model 
to reproduce single and multiple pion 
production rates observed in bubble chamber $\nu_\mu$-deuteron ($A$=2)
and $\nu_\mu$-neon ($A$=20) interactions \cite{Merenyi_92}. 
For pions from vertices placed at random within an iron nucleus,
our model predicts $\sim 45\%$ to emerge either unscattered or 
to have undergone only small angle elastic scattering. About 30\% are
predicted to undergo total absorption, 
while the remainder undergo charge
exchange (9\%) or emerge having undergone inelastic scattering
(16\%).  Our treatment of nuclear rescattering does not include
proton secondaries.  While it is expected that protons would 
occasionally be ejected from parent nuclei 
as the result of rescattering, their momenta would almost always fall
below the effective threshold (approx. 450 MeV/$c$) for creation of 
distinct tracks in the detector.  

   For simulations with intranuclear rescattering included,
the average multiplicity per event for mesons emerging from an
iron nucleus is 3.8 with rms deviation 1.2.  
The average track plus shower multiplicity emitted from an 
iron nucleus (``prong multiplicity") per event is 5.0 with 
an rms deviation of 2.1 prongs. (Note these multiplicities are
for ``perfect detection".)
  
The \nN event generation produces four-momenta of all particles 
which exit the parent nuclei.  This information is fed, 
event-by-event, into the Soudan 2 Monte Carlo (MC) 
which provides a realistic detector response. 
The simulation includes detector background ``noise" arising from
natural radioactivity and from the electronics.  Event records
are generated with format identical to that of data events, allowing
\nn oscillation events to be processed using the same codes and
procedures as for data and for events of the atmospheric
neutrino Monte Carlo.

\subsection{Properties of processed \nN samples}  
\label{sec:processed_samples}

   Three separate $\overline {\rm n}$N simulation samples 
were generated.  For each sample, events were generated 
at random locations throughout the tracking calorimeter as it evolved 
during the experiment's eleven years of data-taking.
In our ``primary" simulation, \nN annihilations were started
at random points throughout their parent nuclei and pion intranuclear
rescattering was implemented.  In a second simulation which also
included intranuclear rescattering, \nN annihilations were 
restricted to the nuclear periphery ($0.75 R \le r \le R$, where $R$ is
the nuclear radius).
A third simulation was carried out in which no pion intranuclear 
rescattering was included.

   Each of the three oscillation samples was processed using
a sequence of selections and procedures very similar to that 
routinely used in reduction of data events in Soudan 2.   
Each sample was subjected (in software) to the
hardware trigger.  Events which passed the hardware trigger
requirements were subjected to two different software ``Filter" codes.  
The Filters impose event containment criteria, e.g. that no track from
the event approaches closer than 20 cm to the detector 
outer surfaces; they also mitigate against backgrounds arising 
from cosmic ray muons, natural radioactivity, and detector noise.
Events which survived the Filters were then subjected to two
separate scans by physicists.  Scanning was carried out 
using interactive color graphics workstations.
The scan rules provided refinements to the Filter selections
and introduced requirements on imaging quality, e.g. an event was 
rejected if ({\it i}) its primary vertex occurred 
in material interior to the detector but not instrumented, 
or if ({\it ii}) its proximity to inter-module gaps 
compromised the reconstruction of the event.

   Events which survived successive application of the 
hardware trigger, software
filters, and physicists' scans are tallied in the upper four rows 
of Table \ref{tbl:survival_rates} for each of the three simulations.
These entries are input to the calculation of the detection
efficiency for neutron-antineutron oscillations in Soudan 2 as will be
described in Sect.~\ref{sec:deteff}.
 
\begin{table}[ht]
\centering
{\normalsize
\begin{tabular}{|l||c|c|c|}
\hline
\it Event Processing and & \it \nN annihilation & \it \nN annihilation in
 & \it \nN annihilation in \\  
\it Selection Stages & \it without INS     & \it nuclear periphery 
 & \it nuclear volume \\ 
 &    &  \it with INS	         &\it with INS           \\
\hline \hline
Initial sample & 491 & 491 & 491 \\ \hline
Hardware trigger   & 490 & 469 & 451 \\ \hline
Containment and quality filters      & 288 & 301 & 286 \\ \hline
Two physicist scans & 214 & 229 & 205 \\ \hline
Multiplicity $\ge$ 4; no proton events & 172 & 148 & 135 \\ \hline
Exclude events with ``muons" & 146 & 130 & 123 \\ \hline
Kinematic selection on $E_{vis}$, $P/E$ & 137 & 102 &  86 \\
\hline
\end{tabular}
}
\caption{Survival through successive processing and selection stages
         for events of three different simulations of \nn oscillations
         yielding \nN annihilations in Soudan 2.}
\label{tbl:survival_rates}
\end{table}

   Preliminary to kinematic reconstruction, the topology of each event 
was characterized in terms of track and shower prongs.  
A track prong results when an ionizing, non-showering charged particle
(e.g. a pion, muon, or proton) traverses drift tubes of
the calorimeter's honeycomb lattice leaving a continuous trail
of tube ``hits".   
A shower prong on the other hand is created by photon conversion or by
a primary electron or positron.  The electromagnetic shower
consists of many distinct particles and exhibits a
cone-like pattern of hits, interspersed with gaps, which is aligned with 
the direction of the initiating photon or electron.   The mean conversion
length for photons in Soudan 2 is approximately 15 cm. 
As a consequence, photon-induced showers generally appear in the vicinity of
but moderately displaced from event primary vertices.

\begin{figure}[hbt]
\centerline{\epsfig{figure=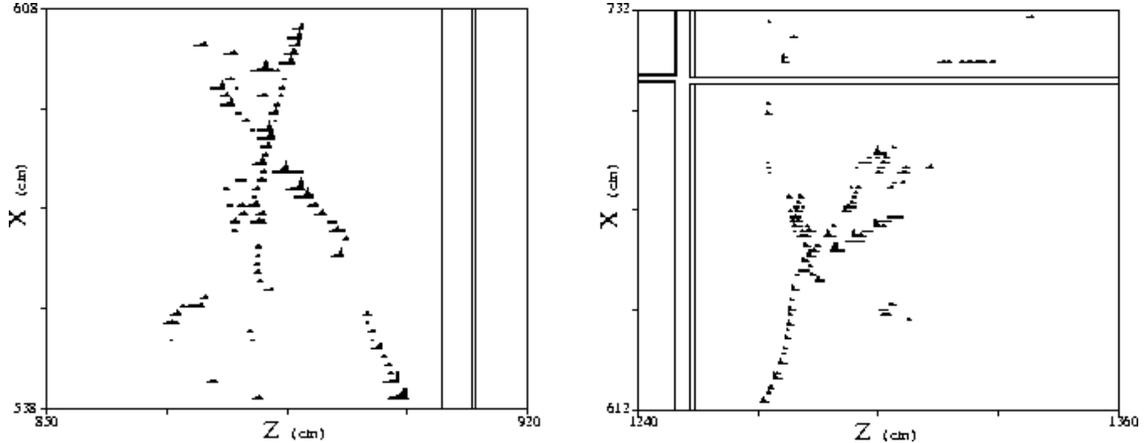,width=15.0cm}}
\caption{Anode versus drift time views (magnified) of two simulated
annihilation events within iron nuclei following \nn oscillation. 
The multiprong hadronic final states of these events are isotropic
to a degree unusual for neutrino interactions.}
\label{fig:nnbmcevents.eps}
\end{figure}

   Images of two \nn oscillation events simulated with full
detector response are displayed in Fig. \ref{fig:nnbmcevents.eps},  
where the anode ($X$) versus drift time ($Z$)
view of each event is shown.  For event scanning, three views
are always used, including cathode-time and anode-cathode as well
as anode-time projections.  (The anode-cathode view is automatically
assembled for all events, MC as well as data, using demultiplexing
and hit-matching algorithms.)
As suggested by Fig. \ref{fig:nnbmcevents.eps}, 
$\overline {\rm n}$N events appear to be energetic
yet isotropic to a degree uncommon for data events and for 
events of the atmospheric neutrino Monte Carlo.

   The occurrence of relatively high multiplicity per event 
of track plus shower prongs is a signature feature of \nN annihilation.
From simulation of \nn oscillation events in the absence of 
intranuclear rescattering, we find the prong multiplicity distribution
- after triggering, filtering, scanning, and reconstruction 
- to have a mean of 6.3 prongs per event with an rms width of 2.8 prongs.
Here the prong multiplicity is relatively high (compared to multiplicities
as generated, see Sect. 2.2), due to depletion of low multiplicities 
by the processing.   
With intranuclear rescattering included, the prong multiplicity
distributions of processed samples are shifted lower.  For both 
simulations with INS included, the multiplicity distributions have
mean values of 4.8 prongs with rms widths of 3.2 prongs.
Thus a sizable contribution from multiplicities 
exceeding three-prongs survives in the latter simulations.  Events of 
this kind are relatively uncommon among contained atmospheric neutrino
events.

\begin{figure}[hbt]
\vspace {320.0pt}
\includegraphics{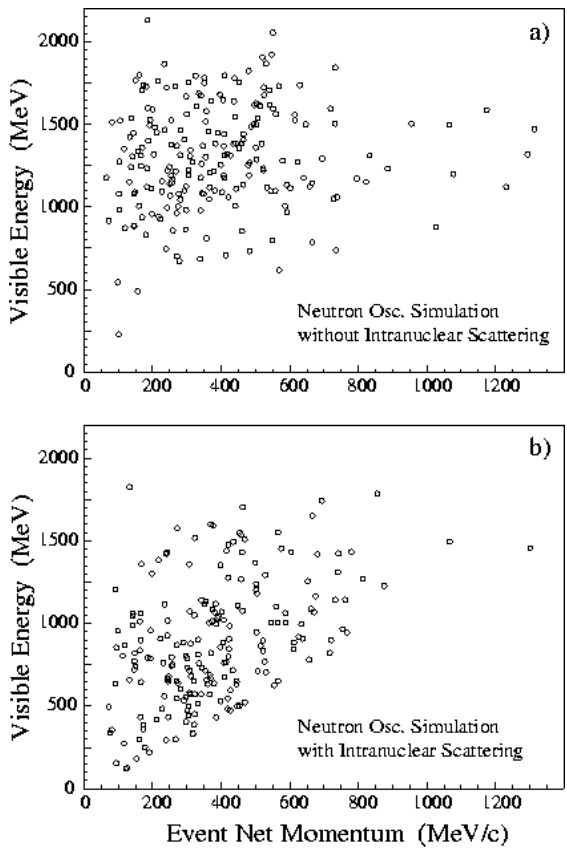}
\caption{Distributions of visible energy versus magnitude of
         vector net momentum for reconstructed \nN annihilation events
         of two independent full detector simulations.  
         The simulation plotted in \ref{fig:nnb_evis_p.eps}a
         (\ref{fig:nnb_evis_p.eps}b)
         omits (includes) intranuclear rescattering of final state 
         pions.  The shift to lower $E_{vis}$ in simulation of b) versus a)
         arises from pion absorption and inelastic scattering within 
         iron nuclei.}
\label{fig:nnb_evis_p.eps}
\end{figure}

   In addition to event topology, the reconstructed kinematic quantities
$E_{vis}$, the visible energy, and $P_{net}$, the net momentum,
provide discrimination between \nN and background events.  These quantities
and their correlation have been used previously in our searches for nucleon
decay \cite{Wall_PRD61_00}.
$E_{vis}$ is calculated as the sum of the relativistic energies
for each of the final state tracks (using the pion mass)
and showers. There occur a few short, heavily ionizing tracks which 
satisfy our identification criteria for protons \cite{Allison_98}.  
For these, only the kinetic energy is added into the calculation of $E_{vis}$.
Distributions of $E_{vis}$ versus $P_{net}$  
for \nN events in the absence of and including 
intranuclear rescattering, are plotted in 
Fig. \ref{fig:nnb_evis_p.eps}a and Fig. \ref{fig:nnb_evis_p.eps}b
respectively.  Reconstructed events, even in the
absence of INS as in Fig. \ref{fig:nnb_evis_p.eps}a, exhibit large 
energy losses arising from low energy prongs and secondary hadronic
scatters which are unresolved.
Consequently the events cluster well below $E_{vis}$ of 1.88 GeV.
Comparison of Fig. \ref{fig:nnb_evis_p.eps}a with Fig. 
\ref{fig:nnb_evis_p.eps}b shows that a further large degradation 
arises with INS.  The INS effects substantially increase 
the kinematic overlap of
\nN events with atmospheric neutrino events.

\section{Data and Backgrounds for the \nN Search}

\subsection{Neutrino, rock, and atmospheric $\nu$ MC events}
\label{sec:nuandrockevts}

   Since \nN events generally have high multiplicities we restrict our
analysis to events having a ``multiprong" topology, that is, having
two or more produced particles emerging from primary vertices.
Quasi-elastic neutrino interactions which produce a charged lepton plus
a recoil proton are readily distinguished and removed.   
As was done for the simulated \nn events, we require
all events under consideration to be fully contained in an interior
volume which is everywhere 20 cm from calorimeter outer surfaces.

   Three multiprong event samples, which are in addition to 
the simulated \nn samples discussed above,
have been isolated for this search:
 
   First and foremost are data events for which the cavern-liner
active shield array was quiescent during the allowed time window.
These events comprise our {\it shield-quiet data sample}.  They
originate mostly with atmospheric neutrino interactions 
but would also include \nn oscillation events.  
There are 188 multiprong events
in the shield-quiet data of this analysis.

   There is a second category of data events which comprises a background
to neutrino multiprongs as well as to \nn oscillations.  Events of this
category are usually produced by energetic neutrons released 
in inelastic cosmic ray muon interactions with the cavern rock 
surrounding the tracking calorimeter.
Most of these {\it rock events} are accompanied by charged
particles which give coincident hits in the
active shield array; they constitute our {\it shield-tagged rock 
event sample}.  However, a few rock events are not
accompanied by shield hits, either because of shield inefficiency 
(detection efficiency is $94\%$) or because no charged particles
enter the cavern along with a rock neutron.   The latter {\it shield-quiet
rock events} may end up as background in shield-quiet data.  
Fortunately the shield-tagged rock events provide a control sample from
which the amount of residual rock background in shield-quiet data may be
inferred.

   The predominant background for \nn oscillations arises from 
the atmospheric neutrino reactions.  Their contribution is evaluated 
using the Soudan 2 {\it atmospheric neutrino Monte Carlo sample}.
Our neutrino MC simulation utilizes the atmospheric flux calculation 
of the Bartol group for the Soudan site \cite{Agrawal_96}. 
Details of the neutrino event generation and comparisons with low
energy $\nu N$ data are given elsewhere \cite{Gallagher_thesis}.
MC atmospheric neutrino events were inserted into the 
experiment's data stream during data-taking.  The MC events were
introduced at a rate 6.06 times higher than the data rate expected
(for null neutrino oscillations) from the atmospheric neutrino flux.  
The events were processed identically to real data events, 
with their identity revealed only at the final analysis stage.  
The atmospheric $\nu$ MC multiprong sample finally extracted 
contains 1267 events.

   Properties of our multiprong data, rock, and neutrino MC samples
germane to an \nn oscillation search are reported below.  
Further details can be found in previous publications 
\cite{Wall_PRD62_00,Wall_PRD61_00,Allison_98}.

\subsection{Multiprong data compared to atmospheric $\nu$ MC}

The prong multiplicity distribution of the shield-quiet data 
sample is summarized using tallies in the grid displayed in
Table \ref{tbl:trk_vs_shw}.  In the Table, the topology of each 
event is represented by the number of tracks, $n_{track}$, 
and number of showers, $n_{shower}$, which comprise the visible
final state.  The number of data events which have a particular
$(n_{track},n_{shower})$ combination are given by the upper entries
at the grid location having the appropriate integer coordinates.
The multiprong data topologies having highest rates are seen to be 
combinations of lowest multiplicity, namely 
$(n_{track}, n_{shower}) = (2,0)$ and (1,1).
For comparison, the corresponding prong distribution 
for the atmospheric neutrino MC event sample 
is given by the lower entries in each of the 
track-shower grid locations of Table \ref{tbl:trk_vs_shw}.
The MC sample (for null neutrino oscillations) 
has been normalized to the same
fiducial exposure as for the data (1267/6.06 = 209.1 events). 
The two distributions are seen 
to match well, e.g. the $\chi^2$/bin averages to below 1.0, with
the exception of the (1,2) topology.
Since most shield-quiet data multiprongs are initiated by
atmospheric neutrinos, the agreement is evidence that
the neutrino MC provides a good general representation of our data.

\begin{table}[ht]
\centering
\begin{tabular}{|c|c|c|c|c|c|c|c|c}
\cline{1-1}
Number of   &\mulc{7}{}{}                 \\ 
Tracks      &\mulc{7}{}{}	  	  \\   \cline{1-8}
\rule[0.4cm]{0mm}{0cm}
\raisebox{-1.0ex}{5}
  &  {\it Data} &  0   & 0    & 0    & 2    & 1    & 1    & \\ 
  &  {\it MC}   & 0.3  & 0.3  & 0.3  & 0   & 0.2 & 0.5 & \\  \cline{1-8}
\raisebox{-1.0ex}{4} 
 &  {\it Data} &  3   & 1    & 3    & 0   & 1   & 0   & \\  
  &  {\it MC}   & 2.1  & 0.3  & 0.2  & 1.6 & 0.2 & 0.2 & \\ \cline{1-8}
\raisebox{-1.0ex}{3}
  &  {\it Data} & 16   & 7    & 2    & 2   & 0   & 0   & \\ 
  &  {\it MC}   & 14.5 & 5.1  & 3.1  & 1.4 & 1.6 & 0.3 & \\ \cline{1-8}
\raisebox{-1.0ex}{2}
  &  {\it Data} & 26   & 20   & 15   & 4   & 0   & 2   & \\ 
  &  {\it MC}   & 26.8 & 21.1 & 14.6 & 5.9 & 2.7 & 1.3 & \\ \cline{1-8}
\raisebox{-1.0ex}{1}
  &  {\it Data} &      & 25   & 10   & 5   & 3   & 3   & \\ 
  &  {\it MC}   &       & 28.3 & 23.6 & 9.3 & 5.0 & 1.6 & \\  \cline{1-8}
\raisebox{-1.0ex}{0}
  &  {\it Data} &      &      & 14   & 13  & 4   & 1   & \\ 
  &  {\it MC}   &      &      & 12.4 & 9.0 & 2.9 & 2.1 & \\  \hline
  &       &      &      &      &     &     &     & \mulc{1}{l|} {Number} \\  
  &       &  0   &  1   & 2    & 3   & 4   & 5   & \mulc{1}{c|} {of} \\  
  &       &      &      &      &     &     &     & \mulc{1}{c|}{Showers}   \\
\hline
\end{tabular}
\caption{Event topology distributions: Event counts are tabulated according to
         $(n_{track},n_{shower})$ combination per event, for multiprong events
         of the data sample (upper entries) and of the atmospheric 
         neutrino MC sample (lower entries).  
         Event tallies of the latter sample represent expectations for    
         an atmospheric flux with null $\numu$ oscillations 
         normalized (factor of 6.06) to the data exposure.}
\label{tbl:trk_vs_shw}
\end{table}

\begin{figure}[hbt]
\vspace {300.0pt}
\includegraphics{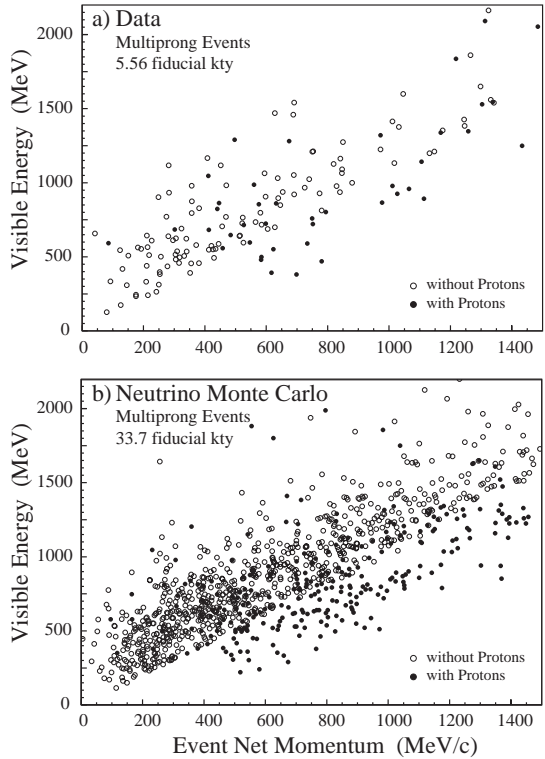}
\caption{$E_{vis}$ versus $P_{net}$ distributions for shield-quiet
         data events and for 
         events of the atmospheric neutrino MC sample.  
         Events without (with) proton 
         tracks are depicted using open (solid) circles.  
         The $\nu$ MC sample (\ref{fig:evis_p_data_numc.eps}a), 
         which corresponds to an
         exposure (for null $\nu$ oscillatiions) of  
         6.06 times the data, reproduces the general 
         kinematic trends of multiprong data 
         (\ref{fig:evis_p_data_numc.eps}b).}
\label{fig:evis_p_data_numc.eps}
\end{figure}

   Kinematics for the shield-quiet multiprong data
is shown by the $E_{vis}$ versus vector net
momentum diplot of Fig. \ref{fig:evis_p_data_numc.eps}a.
The events populate a broad, correlated region extending from threshold
to 1.5 GeV (1.5 GeV/$c$) in visible energy (net momentum).  
Portions of this data distribution overlap the distribution 
predicted for \nn events shown in Fig. \ref{fig:nnb_evis_p.eps}b.

  A direct comparison with kinematics for
the atmospheric neutrino MC sample is provided by
Fig. \ref{fig:evis_p_data_numc.eps}b.  Note that the
latter sample has an exposure-equivalent of 33.7 fiducial kton-years.   
The distribution of the MC sample, normalized to the exposure, is very
similar to that of the data.

  In Fig. \ref{fig:evis_p_data_numc.eps}, events which have
prompt protons are denoted by solid circles.  As previously
noted, such events are highly improbable as \nn oscillations.  
However our atmospheric $\nu $ MC indicates that detectable 
recoil protons are to be
expected in  24\% of neutrino multiprongs, and this expectation
is born out by the frequency of recoil proton events observed in the 
data of Fig. \ref{fig:evis_p_data_numc.eps}a. 
In Fig. \ref{fig:evis_p_data_numc.eps}a there
is a tendency for data events having recoil protons to occur in a region
parallel to and below the more populated kinematic band.
This trend is reproduced in the MC distribution 
of Fig. \ref{fig:evis_p_data_numc.eps}b.

\subsection{Background from rock events}
\label{sec:rockbkgrdindata}

   Rock data events which are flagged by hits in the surrounding
shield array, have a distribution of vertex depth into the detector
which reflects the mean hadronic interaction length of neutrons     
in the Soudan detector medium, approximately 80 cm.
A rock background component in our shield-quiet multiprong
data will have a corresponding vertex depth distribution in the detector.
Neutrino and \nn oscillation events on the other hand, will
distribute uniformly throughout the detector volume.   
Thus by fitting the distribution of vertex depths observed 
in our multiprong data to the sum of two component distributions
representing vertex depths of shield-tagged rock and of neutrino MC
events, the amount of shield-quiet rock contamination 
can be determined \cite{Allison_98}.  On the
basis of a maximum likelihood fit, we estimate that a fraction
$0.06^{+0.11}_{-0.06}$ of shield-quiet data multiprongs 
are produced by rock neutrons.

\section{Selection of \nn Oscillation Events}
\label{sec:Selection_of_Events}
    
   Four selection criteria for neutron oscillation candidates
(Table~\ref{tbl:criteria})
have been defined and applied to all event samples.  The 
selections are designed to maximize \nn detection in the shield-quiet
multiprong data sample while minimizing backgrounds, the principal one
arising from inelastic interactions of atmospheric neutrinos.

\begin{table}[ht]
\centering
{\normalsize
\begin{tabular}{|c|l|}
\hline
\it Cut & \it Definition \\
\hline \hline
1. & $N_{prong} \ge 4$ \\ \hline
2. & No prompt ``proton" tracks \\ \hline
3. & No prompt ``muon" tracks longer than 150 cm \\ \hline
4. & Evt kinematics: $P/E < 0.6, 700 < E_{vis} < 1800$ MeV \\
\hline
\end{tabular}
}
\caption{Selection cuts for \nn oscillation events in Soudan 2.}
\label{tbl:criteria}
\end{table}

\begin{figure}[hbt]
\vspace {300.0pt}
\includegraphics{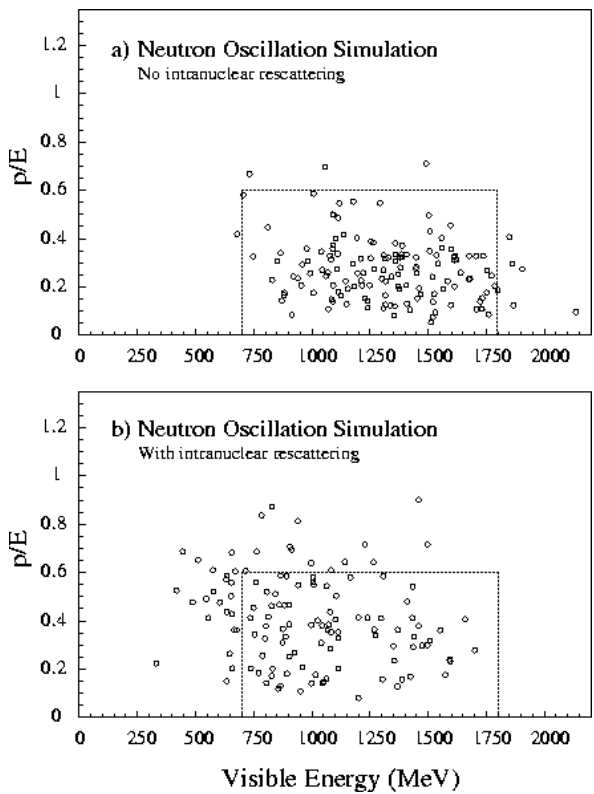}
\caption{Momentum fraction versus visible energy for events from 
         \nn simulation after topology cuts, 
         with intranuclear rescattering omitted 
         (\ref{fig:nnb_sims_final.eps}a) versus included 
         (\ref{fig:nnb_sims_final.eps}b).  The kinematic search
         region of this experiment enclosed by the dotted-line 
         boundary, contains 94\% of events which pass topology 
         selections.  In the more realistic simulation of 
         \ref{fig:nnb_sims_final.eps}b however, INS degrades \evis 
         and increases momentum anisotropy on average,
	 leaving 70\% of events in the search region.}
\label{fig:nnb_sims_final.eps}
\end{figure}

\begin{figure}[hbt]
\vspace {150.0pt}
\includegraphics{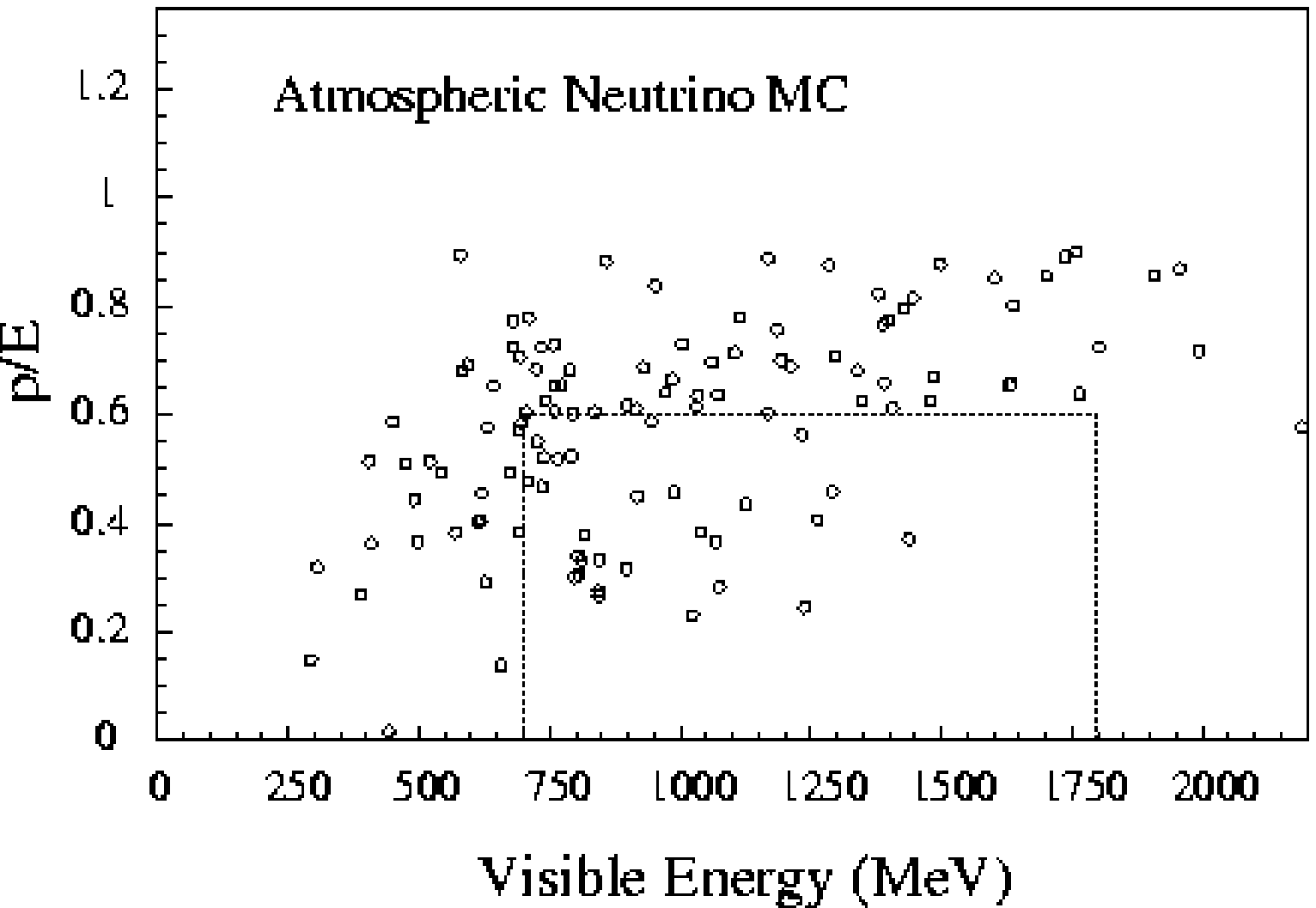}
\caption{Kinematic distribution of atmospheric neutrino MC events 
         which survive topology selections for \nn candidates. Events 
         which occur within the bordered region represent the background 
         expected from an exposure (assuming null neutrino oscillations) 
         which is 6.06 times the data exposure.}
\label{fig:MPMC-FINAL.PS}
\end{figure}

\newcounter{bean}
\begin{list}
{\it\roman{bean})}{\usecounter{bean}
\setlength{\rightmargin}{\leftmargin}}

\item[1.]
{\it Prong multiplicity}: 
We require the sum of track and shower prongs 
directly associated with the primary 
vertices to be greater than or equal to four.  This value is below
but near to the mean multiplicities for processed \nn samples as
discussed in Sect. \ref{sec:processed_samples}. 
Lowering our selection to include three-prongs would increase
the efficiency for \nN detection by only 3\%, however
backgrounds arising from atmospheric neutrinos 
and rock events would increase by 21\% and 32\% respectively.

\item[2.]
{\it Primary proton tracks}:
Events having proton tracks emerging from their primary vertices are
removed from consideration as \nn oscillation candidates.  The rate
of prompt protons is negligible for \nN annihilation events, however
$\sim 24\%$ of atmospheric neutrino events and $\sim 33\%$ of rock
events have visible recoil protons.

\item[3.]
{\it Primary muon tracks}:  Pions produced in \nN annihilations usually
scatter, become absorbed, or range-to-stopping over distances comparable
to the calorimeter's hadronic interaction length of 80 cm.  In order to
mitigate against $\numu$ charged current background, we regard
any non-scattering track which has pion/muon ionization 
and range-to-stopping $\ge$ 150 cm (240 g/cm$^{2}$) to be a muon track.  
Any event having such a track is rejected.  This selection eliminates 36\%
of the atmospheric neutrino sample while cutting less than 10\% of \nn
events of the primary simulation.

\item[4.]
{\it Event kinematics}:
As our final criterion we require the kinematics of
\nn candidates to be compatible with those of \nN annihilation.
The event net momentum fraction $P_{net}/E_{vis}$  
(hereafter $P/E$) has previously been shown to be a useful variable for 
separating \nn oscillation events from 
atmospheric neutrino reactions \cite{Frejus_90}, and its utility
is confirmed in our analysis.
Figures \ref{fig:nnb_sims_final.eps}a and 
\ref{fig:nnb_sims_final.eps}b show $P/E$ versus $E_{vis}$ for \nN events
after topology cuts (1 through 3)
from simulations in which oscillations occur
throughout the parent nuclear volumes.   For the case of no
intranuclear rescattering (Fig. \ref{fig:nnb_sims_final.eps}a), 
approximately 94\% of \nn oscillation events occur in a region 
having $P/E$ less than 0.6, with $E_{vis}$ ranging 
from one to two nucleon masses.  With intranuclear rescattering included, 
Fig. \ref{fig:nnb_sims_final.eps}b, the \nn event 
population shifts to lower values of $E_{vis}$ and to higher
values of $P/E$. Nevertheless a degree of clustering remains which allows a 
useful search region in the ($E_{vis}$, $P/E$) plane to be defined. 
Our selection for \nn kinematics is depicted by 
the rectangular region (dotted-line boundary) shown in  
the Figs \ref{fig:nnb_sims_final.eps}-\ref{fig:MPDATA-FINAL.PS}.  
Candidate \nn events are required to have 
$P/E$ $<$ 0.6 and $700 < \evisn < 1800$ MeV.
These cuts are designed to minimize the atmospheric neutrino 
background shown in Fig. \ref{fig:MPMC-FINAL.PS} 
while maximizing the acceptance for \nN events. 

\end{list}

\section{\nn Candidates and Backgrounds}
\label{sec:deteff}

    The reductions in each \nN annihilation sample upon
successive \nn candidate selections are summarized 
in the bottom three rows of Table \ref{tbl:survival_rates}.  
It can be seen that intranuclear rescattering significantly
lowers the survival rates.
For our primary simulation, with INS operative and with 
annihilations occurring throughout the volume of the parent nucleus,
86 events from an initial 491 event sample survive all selections, 
giving a detection efficiency of 0.18 $\pm$ 0.02.
Detection efficiencies for the comparison simulations are 
higher, $(28 \pm 3)\%$ for the simulation without INS, and
$(21 \pm 2)\%$ for the simulation which includes INS but 
restricts annihilations to the nuclear periphery.

\begin{figure}[hbt]
\vspace {150.0pt}
\includegraphics{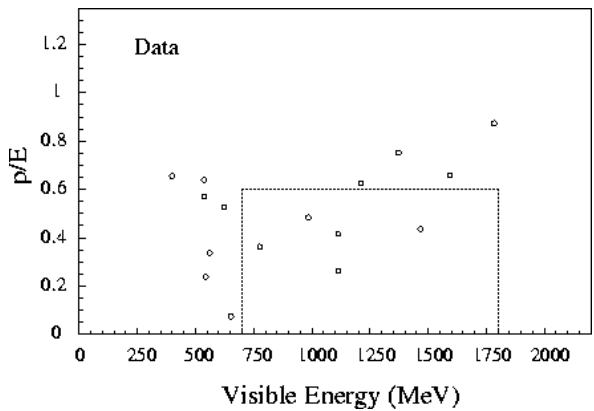}
\caption{Distribution of data multiprong events after 
         topology selections, in the P/\evis versus \evis plane.  
         Five events are observed to have kinematics compatible
         with \nn oscillations in the Soudan 2 detector.}
\label{fig:MPDATA-FINAL.PS}
\end{figure}

   Shield-quiet data events which satisfy \nn selections 1--3 of 
Table~\ref{tbl:criteria}
distribute in the $P/E$ versus \evis plane as shown 
in Fig. \ref{fig:MPDATA-FINAL.PS}.  Of these sixteen events, 
five occur within the kinematically allowed region and are  
candidate \nn oscillation events of this search.   Three of the
candidates have four-prong topologies, while the remaining two are
five-prong events.   

\begin{figure}[hbt]
\vspace {300.0pt}
\includegraphics{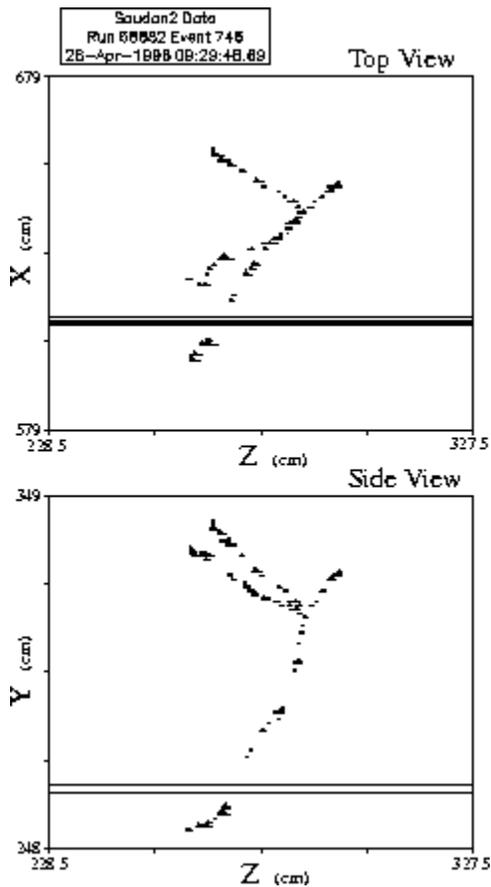}
\caption{Anode-time and cathode-time views of
         a four-track \nn oscillation candidate.}
\label{fig:R68882.PS}
\end{figure}

  Two views of one of the candidates consisting of four pion-or-muon tracks
emerging cleanly from a reaction vertex, 
are shown in Fig. \ref{fig:R68882.PS}.  This event is also consistent with
multipion production by atmospheric neutrinos,
e.g. $^(\bar\nu^) + n \rightarrow \mu^\pm \pi^\mp \pi^+ \pi^- (N)$
\cite{Derrick_84}.

   Fig. \ref{fig:MPMC-FINAL.PS} shows that, among
atmospheric $\nu$ MC events which satisfy topology
selections 1--3 of Table~\ref{tbl:criteria}, there are
28 events (11 $\numu$~CC, 7 $\nue$~CC, and
10 NC) which occur in the selected kinematic region.  
That is, 28 neutrino-induced background events would occur in a
33.7 fiducial kiloton-year exposure to an atmospheric neutrino 
flux having null neutrino oscillations.
However strong evidence for atmospheric neutrino oscillations
now exists \cite{SuperK_OSC}, \cite{MACRO_OSC},
\cite{{Allison_98},{Mann_Sudbury}}.  We allow for neutrino 
oscillations in our background estimate by weighting our
MC $\numu$~CC prediction by the atmospheric neutrino flavor 
ratio-of-ratios:  $R_\nu = [(\numu + \mbox{$\overline{\nu}_{\mu}$})/(\nue + 
\mbox{$\overline{\nu}_{e}$})]_{Data}/[(\numu + \mbox{$\overline{\nu}_{\mu}$})/
(\nue + \mbox{$\overline{\nu}_{e}$})]_{MC}$.
We have used the value of $R_\nu$ measured in this experiment, 
based upon a 5.14 fiducial kton-year exposure: 
$ R_\nu = 0.68 \pm 0.11 \pm 0.06$~\cite{Mann_Sudbury}.
This correction reduces the neutrino background sample 
to 24.5 events.
After scaling to the data exposure and accounting for 
statistical error plus uncertainties arising from our atmospheric neutrino
MC including the neutrino oscillation correction (see 
Sect.~\ref{sec:errors_bkgrds}), we obtain
$4.0 \pm 1.0 \bs $ events as the estimated neutrino-induced 
background of our \nn oscillation search.

   As discussed in Sect.~\ref{sec:nuandrockevts}, neutron-induced 
rock events which are background for \nn oscillations 
are those which elude tagging by the active shield.
As described in Sect.~\ref{sec:rockbkgrdindata}, 
we estimate that a fraction $0.06^{+0.11}_{-0.06}$ of data multiprongs 
are rock events.   Thus of the 188 shield-quiet data multiprongs, 
we estimate $11.3^{+20.6}_{-11.3}$ to be rock events.
However, among the 375 shield-tagged rock multiprong events, 
only fifteen pass our four \nn selection criteria.
Thus, assuming that the selection efficiency of shield-quiet rock events
is the same as that of shield-tagged rock events, the background of rock events
which pass all \nn selections is estimated to be $0.5^{+0.8}_{-0.5}$ events.

   In summary, we estimate the number of background events to be 
$4.0 \pm 1.0 \bs (atm. \bs \nu)$ plus $0.5^{+0.8}_{-0.5} \bs (rock) = 
4.5 \pm 1.2$ events.

\section{Neutron Oscillation Time Lower Limits }

   Having observed five candidate \nn oscillation events and
having calculated background neutrino and cosmic ray processes to
contribute 4.5 events, we set a lower limit 
for the \nn oscillation time $T_A$ in iron nuclei.
We follow previous experiments in using our best estimate of the background 
in determining the limit.  Additional uncertainties in the limit due to the 
uncertainties in the background calculation are discussed in 
Sect. \ref{sec:stat_and_syst_errors}.  We use the approach
to confidence level construction as formulated by Feldman
and Cousins \cite{{Feldman_Cousins},{PDG_00}}. 
For five candidate events with 4.5 background events the
signal limit at 90\% CL is $n_{90} = 5.5$ events.
The oscillation time lower limit is then calculated using

\begin{equation}
\centering
T_{A} > \frac{N_{n} \cdot T_{f} \cdot \epsilon}{n_{90}}   
\label{equ:bound_neutron_limit}
\end{equation}

   Here, $N_n = 3.15 \times 10^{32}$ neutrons in a kiloton 
of the Soudan 2 detector, $T_f = 6.96$ kiloton years 
is the full detector exposure, and $\epsilon$ = 0.18 is
the efficiency for detection of \nn oscillations in the full detector mass, 
calculated using our primary \nn simulation (see Sect.~\ref{sec:deteff}).  
At 90\% CL we obtain $T_{A} > 7.2 \times 10^{31}$ years.

   Equation (\ref{equ:bound_neutron_time}) shows that the free \nn
oscillation time \tnn \bs can be inferred from oscillation time
\TA for neutrons of nucleus $A$. 
For the suppression factor of iron we use the value
$T_{R} = 1.4 \times 10^{23} \bl {\rm s}^{-1}$ 
obtained by Dover, Gal, and Richard upon 
averaging over calculations using different nuclear optical potentials
\cite{Dover_83_85_89}. 
Similar or moderately higher \TR values have been reported from
other calculations\cite{{Huefner_98},{Alberico_91}}.  
At 90\% CL we obtain $\tnn > 1.3 \times 10^{8}$ seconds.

   Oscillation time lower limits based upon
detection efficiencies of each of our three \nn simulations
are summarized in Table \ref{tbl:S2_search_limits}.
The range of values which results from the variation in $\epsilon$
illustrates the important role of the intranuclear rescattering treatment.
The alternative simulations serve as estimators of 
sensitivity to systematic uncertainties arising 
from modelling of the nuclear environment.

\begin{table}
\centering
\begin{tabular}{|l||c|c|} 
\hline
\it \nn Simulation & \it $T_A$ lower limit  & \it \tnn\ lower limit  \\
 \hline \hline
\nn without INS &$  11. \times 10^{31}$ y &$  1.6 \times 10^8$ s  \\ \hline
\nn in Nuclear Periphery, with INS & $  8.4 \times 10^{31}$ y &
 $  1.4 \times 10^8 $ s \\ \hline
\nn throughout Nucleus, with INS &$  7.2 \times 10^{31}$ y 
&$  1.3 \times 10^8 $ s  \\
\hline
\end{tabular}
\caption{Neutron-antineutron oscillation time lower limits (90\% CL) 
         arising from three different treatments of pion 
         intranuclear scattering in \nN annihilations.  Simulations of
         the second and third rows are regarded to be realistic; the 
         one with more conservative nuclear modelling (bottom row) 
         is the basis for the limits for this search.}
\label{tbl:S2_search_limits}
\end{table}

\section{Statistical and Systematic Uncertainties}
\label{sec:stat_and_syst_errors}

   Our lower limits for \TA and \tnn ~contain 
uncertainties arising from finite statistics of data and
simulation samples and from systematic errors inherent in
our analysis techniques.  To quantify these uncertainties, 
we have evaluated the error contributions which enter
via each of the factors of Eq. (\ref{equ:bound_neutron_limit}).

\subsection{ Exposure and \nn detection efficiency}

   The $N_n$ value \cite{S2:NIM_A376_A381} and
the detector exposure are known accurately,
with a consequent fractional error 
for the factor $N_n \times T_f$ of order 2\%.

   The \nn detection efficiency $\epsilon$ 
has statistical error due to the finite sample size of our
primary simulation at generation and especially after cuts.  
Furthermore, systematic uncertainties may arise due to inaccuracies 
in the simulation.  Based upon trial variations of
relative rates among dominant \nN channels (cross sections known to
$\leq 20\%$), we estimate channel rate uncertainties to 
contribute $\sim 5\%$ uncertainty to $\epsilon$.
A systematic error in $\epsilon$ may arise from our placement of 
annihilation sites within parent nuclei.  Based upon our alternative
simulations, we assign a 17\% fractional uncertainty to modelling details.  
There are uncertainties inherent with our treatment of pion 
intranuclear rescattering.  Compared to the unphysical case wherein
INS is neglected, our treatment gives rise to an efficiency decrease
$\Delta \epsilon = 0.10$.  Uncertainty in the amount of this
decrease arises from finite statistics of bubble chamber
neutrino samples \cite{Merenyi_92} and with
extrapolation from A = 20 to 56 of the component INS processes, 
namely pion absorption, charge
exchange, inelastic and elastic scattering.  We estimate the
latter uncertainties to total 30\% \cite{Wall_PRD61_00}.
We then infer the fractional error contribution to $\epsilon$ 
arising from our INS treatment to be $17\%$.  Combining the above
uncertainties in quadrature, we estimate the total fractional
uncertainty $\delta \epsilon/\epsilon$ to be 27\%.

\subsection{ Background estimation }
\label{sec:errors_bkgrds}

   The value of $n_{90}$ relies upon our estimation of rates for 
background events which satisfy the \nn oscillation search criteria.  

   Our estimate of atmospheric neutrino backgrounds is susceptible
to errors from three sources: ({\it i}) Uncertainty arises from
event statistics (after cuts) of the atmospheric 
neutrino simulation (19\%); ({\it ii}) The normalization of the
$\nu$ MC to the experimental exposure, which predicts 183
neutrino events will be observed, is uncertain.  From our data we 
observe (data-rock) = 177 $\pm$ 20 events, from which we infer an
error (11\%). ({\it iii}) The atmospheric $\nu$ MC may not fully
represent aspects of neutrino data which feature in the
selections of Sect. \ref{sec:Selection_of_Events}.  However our data
imply limits on the extent of MC mis-representation.
For data events, and with respect to the cuts applied in 
Sect. \ref{sec:Selection_of_Events},  we observe 34 $\pm$ 5\% to have 
$n_{prongs}$ $\ge$ 4, 78 $\pm$ 9\% to be devoid of protons, 67 $\pm$ 8\%
to be devoid of ``muons", and 43 $\pm$ 6\% to have kinematics in the
vicinity (500 $\le$ $\evis$ $\le$ 1800 MeV, $P/E$ $\le$ 0.8)
of our search region.  The corresponding fractions for the $\nu$ MC
sample are 35\%, 76\%, 64\%, and 39\% respectively.  The agreement
is good, and we take the quadrature sum of the fractional differences
$\mid Data - MC\mid/Data$ in the four ratios as our estimator of uncertainty 
arising from MC representation of atmospheric neutrino physics (11\%). 
The total fractional error for our atmospheric neutrino background 
is then $25\%$, which corresponds to an error of $\pm$ 1.0 events 
on our estimate of 4.0 neutrino-induced background events.  

   Concerning our estimate of background from rock events, the uncertainty 
arising from the determination of rock-event contribution to the vertex depth 
distribution of shield-quiet multiprongs outweighs any other systematic 
uncertainty.  

   The total absolute error assigned to our
4.5 background events from atmospheric neutrinos and rock 
amounts to $\pm$ 1.2 events.
Variation of our background estimate by $\pm$1 $\sigma$ yields
$\delta n_{90}/n_{90} = 21\%$.	The total uncertainty in our
oscillation time lower limit value for \TA thereby implied by 
Eq. (\ref{equ:bound_neutron_limit}) is 
$\delta \TA/\TA = 34\%$.

   Our limit for the oscillation time of free neutrons \tnn
depends upon the nuclear suppression 
factor \TR as implied by Eq. (\ref{equ:bound_neutron_time}).
For \TR of iron, we used the value of Dover \etal \cite{Dover_83_85_89},
$\TRn = 1.4 \times 10^{23} \bl \mb{s}^{-1}$.
However from the range of \TR($A$=56) values indicated by
the calculation of Alberico \etal \cite{Alberico_91},  
a theoretical uncertainty as large as 100\% may be inferred.  
Then the fractional error in \tnn is $\del \tnn/\tnn \approx 53\%.$

   We conclude that the uncertainties $\delta \TAn/\TA$ and $\delta \tnn/\tnn$
on the oscillation time lower limits obtained in this work may be as large
as 34\% and 53\% respectively.  Of course, comparable uncertainties also
apply to other published limits on \nn oscillation times.

\section{Comparison with Previous Experiments}

   Table \ref{tbl:Experiments_and_Results} compares our results with those
obtained by the three most recent of previous searches.   The numbers of
candidate events and corresponding estimates for background rates are
shown in columns four and five. Background estimates for the underground
experiments (1.1, 2.5, 4.5 events for Kamiokande, \frejusn, and Soudan 2 
respectively) are based upon simulations of atmospheric neutrino interactions 
in the detectors.  For the \frejus experiment, an alternative estimate
of 2.1 background events was obtained based upon interactions recorded 
using planar spark chambers exposed to beams of neutrinos and antineutrinos 
at the CERN-PS~\cite{Frejus:NIM_A302}.  
Soudan 2 is the underground experiment with the largest 
exposure.  It is also the only experiment to observe candidate events, although
all of the underground experiments estimated about one background event per
kiloton year.


\begin{table}
\centering
{\normalsize
\tabcolsep 0.01cm
\begin{tabular}{|l|c|c|c|c|c|c|c|} 
\hline
 & \it Source of & \it Exposure & \it Cand.
 & \it Est. & \TR & \TA &\tnn \\
\it Experiments & \it neutrons & \it fiducial (total) & \it Events &
 \it Bkgrd & & & \\
& &kton-years & & &$10^{23}$\ s$^{-1}$ &$10^{31}$\ y &10$^8$\ s \\
\hline \hline
Grenoble('90) \cite{Baldo-Ceolin_90_94}&reactor beam &- &0 &0 &- &- & 0.86 \\ \hline
Kamiokande('86) \cite{KAM_86}&$^{16}$O
 & 1.11 & 0 & 1.2 & 1 & 4.3 & 1.2 \\ \hline
Fr\'{e}jus('90) \cite{Frejus_90}&$^{56}$Fe
 &1.56 (2.56) &0 &2.5,2.1 & 1.4 & 6.5 & 1.2 \\ \hline
Soudan 2 [this study]&$^{56}$Fe
 &5.56 (6.96) &5 &4.5 &1.4 & 7.2 & 1.3 \\ 
\hline
\end{tabular}
}
\caption{Experimental lower limits at 90\% CL for \nn oscillation times
\TA and \tnn\ of bound and free neutrons respectively.  For results 
of this search (bottom row), the background estimate is corrected
for $\nu_\mu \rightarrow \nu_\tau$ oscillations and limits 
are calculated using the Feldman-Cousins method.}
\label{tbl:Experiments_and_Results}
\end{table}

   A degree of caution is warranted when comparing 
oscillation time lower limits for free neutrons as calculated by
underground experiments versus a reactor neutron beam 
experiment (last column of Table \ref{tbl:Experiments_and_Results}). 
The \tnn limits of Soudan 2, \frejusn, and Kamiokande 
are based upon values for the neutron suppression factor 
\TR as calculated by Dover {\it et al.} \cite{Dover_83_85_89}.
However if \TR values for iron and oxygen are considered which are
at the high end of ranges obtained by Alberico {\it et al.}
\cite{Alberico_91}, then limits at or below the Grenoble value
are implied for the underground experiments.

\section{Summary And Conclusion}

   The fine-grained Soudan 2 tracking calorimeter has been used in a
search for neutron-to-antineutron oscillations occurring with bound
neutrons.  Based upon a fiducial exposure of 5.56 kiloton-years of the
underground detector, a new oscillation time lower limit for \nn
oscillations in iron nuclei has been determined at 90\% CL:
\begin{eqnarray}
\TA(Fe) >  7.2 \times 10^{31} \bs \mb{y}.
\end{eqnarray}

   Assuming the suppression factor for iron is
$\TR = 1.4 \times 10^{23} \bs \mb{s}^{-1}$~\cite{Dover_83_85_89},  
the corresponding limit at 90\% CL for n to \n oscillations
of free neutrons is 
\begin{eqnarray}
\tnn > 1.3 \times 10^{8} \bs \mb{s}.
\end{eqnarray}

   The search reported here is background-limited. 
Candidate events are observed to occur at a rate similar to that 
predicted for backgrounds.  Since the predominant background arises
from kinematic overlap with multiprong reactions 
initiated by atmospheric neutrinos, 
it seems unlikely that future underground experiments 
can avoid also becoming background-limited
in larger exposures. 
Thus reactor neutron beam experiments, 
rather than underground experiments monitoring bound neutrons,
may offer a more promising route for future improvements in sensitivity 
to \tnn (see \cite{Kamyshkov_99} and references therein).

\section*{Acknowledgements}

   This work was supported by the U.S. Department of Energy, the U.K. Particle
Physics and Astronomy Research Council, and the State and University of
Minnesota.  We gratefully acknowledge the Minnesota Department of Natural
Resources for allowing us to use the facilities of the Soudan Underground
Mine State Park.


\begin{thebibliography}{99}

\bibitem{Kuzmin_70}  V.A. Kuzmin, JETP Lett. {\bf 12}, 228 (1970).

\bibitem{Sakharov_67} A.D. Sakharov, JETP Lett. {\bf 5}, 24 (1967).

\bibitem{GUTS_non-SUSY} R.N. Mohapatra and R.E. Marshak, 
 Phys. Rev. Lett. {\bf 44}, 1316 (1980); 
 Phys. Lett. {\bf 94B}, 183 (1980);
 L.-N. Chang and N.-P. Chang, Phys. Lett. {\bf 92B}, 103 (1980);
 T.K. Kuo and S. Love, Phys. Rev. Lett. {\bf 45}, 93 (1980);
 R. Cowsik and S. Nussinov, Phys. Lett. {\bf 101B}, 237 (1981);
 S. Rao and R. Shrock, Phys. Lett. {\bf 116B}, 238 (1982); Nucl.
 Phys. {\bf B232}, 143 (1984).

\bibitem{GUTS_SUSY} Z. Chacko and R.N. Mohapatra,
 Phys. Rev. D {\bf 59}, 055004 (1999);
 K. S. Babu and R. N. Mohapatra, 
 Phys. Lett. B {\bf 518}, 269 (2001).

\bibitem{GUTS_higherD} Stephan J. Huber and Qaisar Shafi,
 Phys. Lett. B {\bf 512}, 365 (2001); S. Nussinov and R. Shrock,
 Phys. Rev. Lett {\bf 88}, 171601 (2002). 

\bibitem{ORNL-6910_Alberico} W. M. Alberico, {\it in:}
 Proceedings of the Workshop on Future Prospects 
 of Baryon Instability Search in p-Decay and N-\={N} Oscillation Experiments, 
 ed. S.J. Ball and Y. A. Kamyshkov, ORNL-6910, March 1996; pp. 221-234.

\bibitem{Dover_83_85_89} C.B. Dover, A. Gal and J.M. Richard, 
 Phys. Rev. D {\bf 27}, 1090 (1983); Phys. Rev. C {\bf 31}, 1423 (1985);
 Nucl. Instr. Meth. A {\bf 284}, 13 (1989). 

\bibitem{Huefner_98} J. Huefner and B.Z. Kopeliovich,
 Mod. Phys. Lett. A {\bf 13}, 2385 (1998).

\bibitem{Chetyrkin_81} K.G. Chetyrkin, M.V. Kazarnovsky, V.A. Kuzmin, and 
 M.E. Shaposhnikov, Phys. Lett. {\bf 99B}, 358 (1981).

\bibitem{Alberico_82_84} W.M. Alberico, A. Bottino and A. Molinari, Phys. Lett. 
 {\bf 114}B, 266 (1982);
 W.M. Alberico, J. Bernabeu, A. Bottino and A. Molinari, 
 Nucl. Phys. A {\bf 429}, 445 (1984).

\bibitem{Alberico_91} W.M. Alberico, A. De Pace and M. Pignone, Nucl. Phys. 
 A {\bf 523}, 488 (1991).

\bibitem{Kondratyuk_96} L.A. Kondratyuk, JETP Lett. {\bf 64}, 495 (1996).

\bibitem{Bressi_89} G. Bressi {\em et al.}, Z. Phys. C {\bf 43}, 175 (1989).

\bibitem{Fidecaro_85} G. Fidecaro {\em et al.}, Phys. Lett. {\bf 156}B, 
 122 (1985).

\bibitem{Baldo-Ceolin_90_94} M. Baldo-Ceolin {\em et al.},
 Phys. Lett. B {\bf 236}, 95 (1990); Z. Phys. C {\bf 63}, 409 (1994).

\bibitem{Homestake_83} Homestake Collaboration, M. L. Cherry {\em et al.},
 Phys. Rev. Lett. {\bf 50}, 1354 (1983).

\bibitem{NUSEX_83} NUSEX Collaboration, G. Battistoni {\em et al.},
 Phys. Lett. B {\bf 133}, 454 (1983).

\bibitem{KOLAR_86} KOLAR Collaboration, M.R. Krishnaswamy {\em et al.}, 
 Nuovo Cim. C {\bf 9}, 167 (1986).

\bibitem{IMB_84} IMB Collaboration, T.W. Jones {\em et al.}, Phys. Rev. Lett. 
 {\bf 52}, 720 (1984).

\bibitem{KAM_86} Kamiokande Collaboration, M. Takita {\em et al.},
 Phys. Rev. D {\bf 34}, 902 (1986).

\bibitem{Frejus_90} Fr\'{e}jus Collaboration, Ch. Berger {\it et al.},
 Phys. Lett. B {\bf 240}, 237 (1990).

\bibitem{S2:NIM_A376_A381} Soudan 2 Collaboration, W.W.M. Allison 
 {\it et al.}, Nucl. Instr. Meth. A {\bf 376}, 36 (1996);
  Nucl. Instr. Meth. A {\bf 381}, 385 (1996).

\bibitem{S2:NIM_A276} Soudan 2 Collaboration, W.P. Oliver 
 {\it et al.}, Nucl. Instrum. Methods Phys. Res. A {\bf 276}, 371 (1989).

\bibitem{Wall_PRD62_00} Soudan 2 Collaboration, D. Wall {\it et al.},
 Phys. Rev. D {\bf 62}, 092003 (2000).

\bibitem{Bodek_81} A. Bodek and J.L. Ritchie, Phys. Rev. D {\bf 23}, 
 1070 (1981).

\bibitem{Chaffee_SAGE}  R.B. Chaffee, SLAC Computation Group Technical Memo
 No. 195, 1979.

\bibitem{Backenstoss_83} G. Backenstoss {\it et al.}, Nucl. Phys. B {\bf 228},
 424 (1983).  A convenient compilation of channel rates based on this work
 appears in Yu.A. Golubkov and M.Yu. Khlopov, 
 astro-ph/0005419; submitted to Yad. Fiz.

\bibitem{pbar_p_Refs} N. Horwitz {\it et al.}, Phys. Rev. {\bf 115}, 
 472 (1959); R. Armenteros and B. French, {\it in:}
 High Energy Physics, edited by E.H.S. Burhop 
 (Academic, New York, 1969), Vol. 4, p. 237;
 C. Ghesquiere, {\it in:} Proceedings of 
 the Symposium on Antinucleon-nucleon Interactions, 
 Liblice-Prague, Edited by L. Montanet, CERN 74-18, (1974), p. 436;
 P. Pavlopoulous {\it et al.}, {\it in:} 
 Nucleon-Nucleon Interactions 1977, Proc. of the Second Int. Conf., 
 Vancouver,  Edited by H. Fearing, D. Measday, and A. Strathdee 
 (AIP Conf. Proc. No. 41), (AIP, New York, 1978), p. 340.

\bibitem{Merenyi_92} R. Merenyi {\it et al.}, Phys. Rev D {\bf 45}, 743 (1992).

\bibitem{Wall_PRD61_00} Soudan 2 Collaboration, D. Wall {\it et al.},
 Phys. Rev. D {\bf 61}, 072004 (2000). 

\bibitem{Allison_98} Soudan 2 Collaboration, W.W.M. Allison {\it et al.},
 Phys. Lett. B {\bf 427}, 217 (1998); see Sect. 3.2.

\bibitem{Agrawal_96} V. Agrawal, T.K. Gaisser, P. Lipari, and T. Stanev,
 Phys. Rev. D {\bf 53}, 1313 (1996).

\bibitem{Gallagher_thesis} H.R. Gallagher, Ph.D. thesis, 
 University of Minnesota, 1996; H. Gallagher and M. Goodman, Soudan 2
 internal note PDK-626 (MINOS note NuMI-L-112), 1995(unpublished). 

\bibitem{Derrick_84} M. Derrick {\it et al.}, Phys. Rev. D {\bf 30},
 1605 (1984).

\bibitem{SuperK_OSC} Super-Kamiokande Collaboration, Y. Fukuda 
 {\it et al.}, Phys. Lett. B {\bf 433}, 9 (1998); Phys. Rev. Lett. {\bf 81},
 1562 (1998); Phys. Lett. B {\bf 436}, 33 (1998); Phys. Rev. Lett. {\bf 85},
 3999 (2000).

\bibitem{MACRO_OSC} MACRO Collaboration, S. Ahlen {\it et al.}, Phys. Lett.
 B {\bf 357}, 481 (1995); M. Ambrosio {\it al.}, Phys. Lett. B {\bf 434},
 451 (1998); Phys. Lett. B {\bf 478}, 5(2000); Phys. Lett. B {\bf 517},
 59 (2001). 

\bibitem{Mann_Sudbury} Soudan 2 Collaboration, W.A. Mann,  
 Nucl. Phys. B (Proc. Suppl.) {\bf 91}, 134 (2001).

\bibitem{Feldman_Cousins} G. J. Feldman and R. D. Cousins,
 Phys. Rev. D {\bf 57}, 3873 (1998).

\bibitem{PDG_00} Particle Data Group, Eur. Phys. J. C {\bf 15}, 201 (2000).


\bibitem{Frejus:NIM_A302} Fr\'{e}jus Collaboration, Ch. Berger {\it et al.},
 Nucl. Instr. Meth. A {\bf 302}, 406 (1991).

\bibitem{Kamyshkov_99} Yu. Kamyshkov,
``Nucleon Instability and ($B-L$) Non-Conservation",
 {\it in:} Proc. NNN99 Workshop, AIP Conf. Proc. 533, Edited by M.V. Diwan 
 and C.K. Jung, Stony Brook, New York, September 1999; p. 84.

\end{thebibliography}
\end{document}